\documentstyle[10pt,epsfig]{article}
\begin{document}
\textwidth 6.75in
\textheight 8.5in

\begin {center}
{\Large A glueball component in $f_0(1790)$}

\vskip 5mm
{D.V.~Bugg\footnote{email address: D.Bugg@rl.ac.uk}},   \\
{Queen Mary, University of London, London E1\,4NS, UK}
\end {center}
\vskip 2.5mm

\begin{abstract}
The $\omega \phi$ threshold peak observed by BES in
$J/\Psi \to \gamma (\omega \phi )$ may be interpreted
quantitatively in terms of a glueball component in $f_0(1790)$.
\end{abstract}

\vskip 2mm

The BES collaboration has provided strong evidence for an
$\omega \phi$ peak at threshold in $J/\Psi \to \gamma (\omega \phi )$
with spin-parity $J^P = 0^+$ [1].
Bicudo et al. argue that this may be interpreted in terms of a $0^+ $
glueball at $\sim 1810$ MeV [2].
Here, their argument is developed quantitatively.
The conclusion is that the $\omega \phi$ signal originates from the
$f_0(1790)$ observed by BES in $J/\Psi \to \phi (\pi ^+\pi ^-)$ [3].
However, the signal appears to be too weak for $f_0(1790)$ to be a pure
glueball.
Instead, the natural explanation is that the glueball is distributed
between $f_0(1370)$, $f_0(1500)$, $f_0(1710)$ and $f_0(1790)$, with
$f_0(1790)$ having a component of $\sim 40\%$ in intensity.

A glueball is a flavour singlet.
In its decay, each gluon converts into a $^3P_0$ combination
$(u\bar u + d\bar d + s\bar s)$.
The final state has flavour content
\begin {equation}
F = (u\bar u + d\bar d + s\bar s)(u\bar u + d\bar d + s\bar s).
\end {equation}
If the decay is to vector mesons, the component $(u\bar u + d\bar d)
(u\bar u + d\bar d)$ makes three charge combinations of $\rho \rho$
and one of $\omega \omega$.
The component $2(u\bar u + d\bar d)s\bar s$ can make $4\omega \phi$
or $2(K^{*0}\bar K^{*0} + K^{*+}K^{*-})$ or some linear combination.

There are BES I data on $J/\Psi \to \gamma
(K^+\pi ^- K^-\pi ^+)$  showing that the channel  $\gamma (K^*\bar K^*)$
does not contain any significant $0^+$ signal [4].
The paper says: `Contributions from $0^{++}$ and $4^{++}$ are small
or absent'.
The branching fraction reported for $J/\Psi \to \gamma (K^*\bar K^*)$
is $4.0 \times 10^{-3}$.
A signal with the same magnitude as that of  $J/\Psi \to \gamma (\omega
\phi)$ in Ref. [1] would be rather conspicuous near 1800 MeV,
because of the small phase space at that mass for $K^*\bar K^*$.

If the glueball component goes preferentially to $\omega \phi$,
the branching ratio $BR[J/\Psi \to \gamma (\omega \phi )]/
BR[J/\Psi \to \gamma (\rho \rho )]$ should be 4/3.
An analysis of Mark III data on $J/\Psi \to \gamma (\pi ^+\pi ^- \pi ^+
\pi ^-)$ reports three $0^+$ peaks at 1500, 1750 and 2100 MeV [5].
The branching fraction reported for $J/\Psi \to \gamma f_0(1750) \to
\gamma (\rho \rho )$ is $[1.9 \pm 0.14(stat) \pm 0.28(syst)] \times
10^{-4}$.
This agrees with $3/4$ times the branching fraction
$2.61 \pm 0.27 (stat) \pm 0.65 (syst)$
reported by BES for $J/\Psi \to \gamma (\omega \phi)$.

Why should the glueball prefer to decay to $\omega \phi$ rather than
$K^*\bar K^*$?
The following argument rationalises the experimental facts.

It is well known that resonances tend to lock to sharp thresholds.
Classic examples are $f_0(980)$ and $a_0(980)$ at the
$K\bar K$ threshold.
The mechanism of the locking process is highly non-linear and
analogous to the operation of the phase-locked loop in a mobile phone.
For details of the electronic case, see the textbook of Best [6].
The vital points will be outlined here and related to the particle
physics case.

In a mobile phone, an oscillator scans a range of frequencies.
Strong non-linearity in the detector generates a beat
frequency between the oscillator and the incoming signal.
A low frequency filter separates out the beat frequency.
The oscillator frequency is controlled by a feedback loop which
dissipates the beat signal and locks to the incoming signal.
There are many incoming frequencies, but the system locks to the one
with the sharpest signal, i.e. the lowest range of frequencies.

The $\omega \phi$ channel has a sharp threshold at 1801 MeV.
The amplitude for $\omega \phi$ elastic scattering has a
scattering length with an imaginary component proportional
to the probability of de-excitation to all open channels.
The step in the imaginary part of the amplitude at threshold
produces a sharp peak in the real part of the amplitude,
via analyticity.

\begin{figure}      [htb]
\begin{center}
\epsfig{file=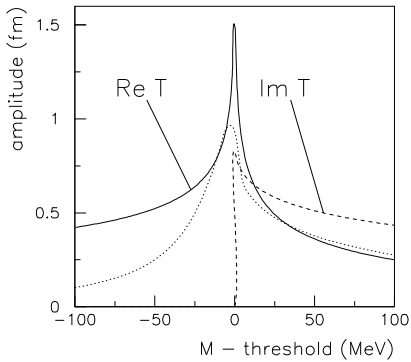,width=7cm}\
\caption{$\mathrm {Re}\, T$ (full curve),
$\mathrm {Im}\, T$ (dashed) calculated from the threshold step,
compared with the actual line-shape of $f_0(980)$ (dotted). }
\end{center}
\end{figure}
An illustration of such a peak is given in Fig. 1 for the case of
$f_0(980)$, whose parameters are known.
The step in the imaginary part of the elastic $K\bar K$ amplitude
is calculated from $g^2(K\bar K)$ of Ref. [3] and
is shown by the dashed curve. The full curve shows the
corresponding peak in the real part of the amplitude.
This peak provides additional attraction at the threshold.
Suppose that in the absence of this effect, the resonance would lie
close to the threshold. The Breit-Wigner amplitude $T$ has the form
\begin {eqnarray}
T(K\bar K \to K\bar K) &=& \frac {g^2_{K\bar K}\rho _{K\bar K}}
{M^2 - s - m(s) - iM\Gamma (s)} \\
m(s) &=& \frac {1}{\pi}\int \frac {Im~M\Gamma (s')ds'}{s' - s},
\end {eqnarray} where $g$ is the coupling
constant to the $K\bar K$ channel and $\rho$ is phase space.
The dispersive term $m(s)$ enhances the resonance near threshold.
The peak in the real part of the amplitude introduces a
phase variation which pulls the resonance towards the threshold.
It settles there with a width which is broadened by the decay
of the resonance.
This line-shape is illustrated by the dotted curve in Fig. 1, using
current parameters of $f_0(980)$.
A similar study of the locking of $a_0(980)  $ to the $K\bar K$
threshold has been made by Rupp and van Beveren [7].
They show that the resonance lies close to the threshold for a
wide range of coupling constants, gradually changing from a
virtual state to a bound state as the coupling to the $KK$ channel
increases.

The $f_0(1790)$ will have an analogous line-shape induced by
decays to $\omega \phi$.
The threshold provides the non-linearity and the
peak in the real part of the amplitude plays the role of the
low-pass filter.
The $K^*\bar K^*$ channel may in principle act in the same way.
However, because of the width of the $K^*$'s, the non-linearity
is much smaller and the filter is much wider.
As a result, the dissipation of the resonance through the
$K^*\bar K^*$ channel is likely to be much less than through the
$\omega \phi$ channel.

Let us now turn to the question whether
the $f_0(1790)$ is likely to be a pure $0^+$ glueball.
Close, Farrar and Li have used sum rules for $J/\Psi \to \gamma (gg)$
(where $g$ is a gluon) to predict branching fractions for glueball
production [8].
The branching fraction depends on the mass and width of $f_0(1790)$.
For the observed mass and width, $\Gamma = 270 ^{+60}_{-30}$ MeV [3],
the predicted branching fraction for $J/\Psi \to \gamma f_0(1790)$ is
$3.4 \times 10 ^{-3}$ if it is a pure glueball.
Observed  branching fractions are collected into Table 1.

\begin{table}[htb]
\begin {center}
\begin{tabular}{cccc}
\hline
Decay channel  & branching fraction $\times 10^4$ & reference \\\hline
$\omega \omega + \rho \rho$ & $2.5 \pm 0.18 \pm 0.38$ & [5]\\
$\omega \phi$ & $2.61 \pm 0.27 \pm 0.65$ & [1] \\
$\sigma \sigma $ & $9.0 \pm 1.3$ & [5] \\
$\pi \pi$ & $1.5 \pm 0.21$ & [9] \\
$K\bar K$ & $\sim 0.5$ & [3] \\
total & $16.1 \pm 1.6$ \\\hline
Prediction & $34 ^{+8}_{-4}$ & [8]\\
$q\bar q$ & $\sim 7$ & [8] \\\hline
\end{tabular}
\caption{Branching fraction for production of $f_0(1790)$ in
$J/\Psi $ radiative decays. }
\end {center}
\end{table}

Entries 1 and 3 come from $J/\Psi \to \gamma (\pi ^+\pi ^-\pi ^+\pi
^-)$ data after correcting for other charge states.
Entry 4  requires discussion of a small minefield of problems.
The $f_0(1790)$ was observed by BES in $J/\Psi \to \phi (\pi ^+\pi ^-)$,
so it clearly requires $\pi \pi$ decays.
However, one can place some limit on the $\pi \pi$ branching ratio
from two sources.
The first concerns DM2 data on $J/\Psi \to \gamma (\pi ^+\pi ^-)$
[9].
They observe a possible signal attributed to
$J/\Psi \to \gamma f_2(1720)$
with a branching fraction of $(1.5 \pm 0.24 \pm 0.23) \times 10^{-4}$,
after including the $\pi ^0 \pi ^0$ contribution.
Today, it is generally agreed that $f_J(1720)$ has spin 0, but
this will only affect the branching fraction by a small amount.
The next point is that BES put an upper limit of 11\% with 95\%
confidence on the ratio $BR[f_0(1710) \to \pi \pi]/[BR[f_0(1710) \to
K\bar K]$ [3].
This implies that any signal observed in $\pi \pi$ by
DM2 comes from $f_0(1790)$ instead of $f_0(1710)$; however, because
the DM2 signal is at 1720 MeV rather than 1790, it may be an upper
limit.

The second source of information is that $f_0(1790)$ is not observed in
Cern-Munich data for $\pi \pi$ elastic scattering [10].
A re-analysis of those data limits its branching ratio to
$\Gamma _{2\pi }/\Gamma _{total} < 0.1 [11]$.
If decays to $4\pi$,
$\omega \phi$, $\pi \pi$ and $KK$ account for all decays of
$f_0(1790)$, this places an upper limit of $1.5 \times 10^{-4}$ on
entry 4 of the Table, like that of DM2.

Entry 5 is obtained from entry 4 using the $K\bar K/\pi \pi$
branching ratio reported by BES for $f_0(1790)$ [3].
Finally, there is evidence for an $\eta \eta$ signal due to
either or both of $f_0(1710)$ and $f_0(1790)$, but with a mass
of $1770 \pm 12$ MeV and a width of $220 \pm 40$ MeV, which
are closer to $f_0(1790)$ than $f_0(1710)$ [12];
however, its branching ratio is much smaller than to $\pi \pi$
and hence negligible.

There is one further point.
$J/\Psi$ radiative decays certainly produce some well known $q\bar q$
states, e.g. $f_2(1270)$. There must likewise be some contribution to
production of a $q\bar q$ component of $f_0(1790)$.
The $f_2(1270)$ signal observed in $J/\Psi $ radiative decays has a
branching fraction of $(6.86 \pm 0.27 \pm 1.03) \times 10^{-4}$ in
decays to $\pi ^+\pi ^-$.
This number needs to be multiplied by 3/2 to allow for $\pi ^0
\pi ^0$ decays, but it also needs to be divided by the same factor for
the relative number of partial waves for $J/\Psi \to 2^+$ and $0^+$.
The $q\bar q$ component needs to be added to the prediction of
Close, Farrar and Li.

In summary, it looks unlikely that the total branching fraction
of $f_0(1790)$ in $J/\Psi $ radiative decays is sufficient to
agree with the prediction of Close, Farrar and Li.

The conventional view, advanced in Refs. [13, 14, 15], is that
there is one $f_0$ too many to be explained as $q\bar q$
in the mass range 1300-1700 MeV.
The $f_0(1370)$, $f_0(1500)$ and $f_0(1710)$ are taken as mixed
states of $n\bar n$, $s\bar s$ and $gg$.
This mixing scheme now needs to be extended to $f_0(1790)$.

The resulting glueball component of $f_0(1790)$ is $\sim 40\%$ in
intensity.
This is rather high.
One may speculate that a `dressed' gluon has a
mass of order 700--800 MeV, and therefore a small radius;
if the glueball is correspondingly compact, its wave function will
overlap well with a $q\bar q$ radial excitation having a node in its
radial wave function.

Contributions to the outstanding glueball and $n\bar n$ components are
$(1.03 \pm 0.14) \times 10^{-3}$ from $J/\Psi \to \gamma
f_0(1500)$ [16] and $8.5 ^{+1.2}_{-0.9} \times 10^{-4}$ from
$J/\Psi \to \gamma f_0(1710)$ [17].
There is a slight short-fall compared with prediction, but it is
not clear how to allow for dependence on mass and width after summing
over several resonances.

In conclusion, the BES data on $J/\psi \to \gamma (\omega \phi )$
provide a new type of input to the discussion of the $0^+$
glueball.
The $f_0(1790)$ is readily accomodated as the
radial excitation of $f_0(1370)$, but with a rather large
glueball component.
This component has the potential to decay to
$\omega \phi$ or $K^*\bar K^*$, but the latter is
observed experimentally to be small.
The $\omega \phi$ component to which the glueball can decay
explains naturally the BES observation of $J/\Psi \to \omega \phi$ at
threshold.
The observed branching fraction for $J/\Psi \to \gamma (\omega \phi )$
agrees with the expected (4/3) times that for $J/\Psi \to
\gamma f_0(1790)$, $f_0(1790) \to \rho \rho$.

\end {document}